\documentclass{aa}
\usepackage{txfonts}

\usepackage[authoryear]{natbib}
\usepackage{graphicx}
\usepackage{txfonts}

\def\degr{\hbox{$^\circ$}}

\def\arcsec{\hbox{$^{\prime\prime}$}}
\def\oIa{\ion{[O}{I]}63\,$\mu$m}
\def\oIb{\ion{[O}{I]}145\,$\mu$m}
\def\oIc{\ion{[O}{I]}63 and 145\,$\mu$m}
\def\cII{\ion{[C}{II]}158\,$\mu$m}
\def\nII{\ion{[N}{II]}122\,$\mu$m}

\begin{document}

\title{The Spatial Variation of the Cooling Lines in the Reflection
  Nebula NGC\,7023
\thanks{{\it Herschel} is an ESA space observatory with science
instruments provided by European-led Principal investigator consortia and with important participation from NASA.}}

\author{J. Bernard-Salas\inst{1,2} 
\and E. Habart\inst{2}
\and M. K\"ohler\inst{2} 
\and A. Abergel\inst{2}
\and H. Arab\inst{2,3}
\and V. Lebouteiller\inst{4} 
\and C. Pinto\inst{5}
\and M.H.D. van der Wiel\inst{6}
\and G.J. White\inst{1,7} 
\and M. Hoffmann\inst{2,8}
}

\offprints{J. Bernard-Salas, \email{jeronimo.bernard-salas@open.ac.uk}}

\institute{Department of Physical Sciences, The Open University, MK7 6AA, Milton Keynes, UK 
\and Institut d'Astrophysique Spatiale, Paris-Sud 11, 91405
  Orsay, France
\and Space Telescope Science Institute, 3700 San Martin Drive, Baltimore, MD 21218, USA 
\and Laboratoire AIM, CEA/DSM-CNRS-Universite Paris Diderot, DAPNIA/Service d'Astrophysique, Saclay, France
\and Laboratoire d'Astrophysique de Marseille, CNRS/INSU Universit\'e de Provence, 13388 Marseille, France
\and Institute for Space Imaging Science, Department of Physics \&
Astronomy, University of Lethbridge, Lethbridge, AB T1K\,3M4, Canada
\and Space Science \& Technology Division, The Rutherford Appleton  Laboratory, Chilton, Didcot OX11 0NL, UK
\and Wolfson Brain Imaging Centre, Addenbrooke\'s Hospital, University of 
Cambridge, Cambridge, CB2 0QQ, UK
}

\date{Received date / Accepted date}

\abstract {The north-west photo-dissociation
  region (PDR) in the reflection nebula NGC\,7023 displays a complex structure. Filament-like condensations at the edge 
  of the cloud can be traced via the emission of the main cooling lines, offering a great opportunity
   to study the link between the morphology and  energetics of these
  regions.}
{We study  the spatial variation of the far-infrared fine-structure lines of
  \ion{[C}{II]} (158\,$\mu$m), \ion{[O}{I]} (63 and 145\,$\mu$m). These lines trace the local gas conditions across the PDR. We also compare their emission to molecular
tracers including rotational and rovibrational H$_2$, and high-rotational lines of CO.} 
{We use observations from the Herschel/PACS instrument  to map the spatial distribution
  of these fine-structure lines.  The observed region covers a square area of
  about 110$\arcsec$x110$\arcsec$ with an angular
  resolution that varies from 4\arcsec~to 11\arcsec. We compare this emission  to 
ground-based and Spitzer observations of H$_2$, Herschel/SPIRE observations of CO lines, and Spitzer/IRAC 3.6\,$\mu$m 
images that trace the emission of polycyclic aromatic hydrocarbons. We use a PDR code to  model
 the \ion{[O}{I]}145$\mu$m line,   
and infer the physical conditions in the region.}
{ The \ion{[C}{II]} (158\,$\mu$m) and \ion{[O}{I]} (63 and 145\,$\mu$m)  lines arise from the warm cloud surface where the
   PDR is located and the gas is warm, cooling the region. 
We find that although the relative contribution to the cooling budget over the observed region is dominated  by \oIa ($>$30\%),  H$_2$ contributes significantly in the PDR ($\sim$35\%), as does \cII~outside the PDR~(30\%). Other species contribute little to the cooling (\oIb~ 9\%, and CO 4\%).   
   Enhanced emission of these far-infrared atomic lines
    trace the presence of condensations, where high excitation CO rotational lines and dust emission 
    in the submillimitre are also detected. The \ion{[O}{I]} maps resolve these condensations into two structures, and show that 
    the peak of \ion{[O}{I]} is slightly displaced from the molecular H$_2$ emission. The size of these structures is around 8$\arcsec$ (0.015~pc), and in surface cover about 9\% of the PDR emission.
We have tested whether the 
   density profile and peak densities derived in previous studies to model the dust and molecular emission, can predict the \oIb~emission.  We find that the model with a peak density of 10$^{6}$ cm$^{-3}$, and 2$\times$10$^{4-5}$ cm$^{-3}$ in the oxygen emitting region, predicts an \oIb~line which is just 30\% less 
   than the observed emission. Finally, we have not detected emission from 
   \ion{[N}{II]}122\,$\mu$m, suggesting that the cavity is mostly filled  with non-ionised gas. } 
 {}


\authorrunning{Bernard-Salas et al.}
\titlerunning{Cooling lines in NGC\,7023}  

\maketitle

\section{Introduction}

NGC\,7023 is a bright and well-known reflection nebula. Although it is a  relatively nearby object, its reported distance  varies significantly in the literature. Using  Hipparcos parallaxes \citet{van97} and later on \citet{lee07} measured a distance of 430$^{+160}_{-90}$
pc and 520$\pm$180~pc respectively, and \citet{ben13} estimated a distance of 320$\pm$51 pc based on revised orbital parameters and astrometry. In this paper we adopt the distance given by \citet{van97} (430~pc), which is almost the  average of the values given by \citet{lee07} and \citet{ben13}.

It is excited by the Herbig Be binary star system HD\,200775 \citep{ale08}, which is immersed
in the cavity of the molecular cloud \citep{fue92}. The overall morphology of NGC\,7023 is driven by the action of the ionising stars, where the stellar outflow, now inactive, creates a biconical cavity  in the shape of a butterfly of about of 1.5~pc$\times$ 0.8~pc in size \citep{koh14}.  The stars illuminate the walls of the cavity in an almost edge-on orientation.  At the edge of these walls, three PDRs are present \citep{fue97}, and are located at 40$\arcsec$ northwest
(herein NW), 70$\arcsec$ south-west (SW), and 170$\arcsec$ east of the stars. The NW PDR is
the brightest, and is oriented almost edge-on to the
observer. Polarization observations in the near-infrared \citep{sel92}, as well as vibrational H$_2$ \citep{lem96} of the NW PDR reveal a filamentary morphology.  \citet{wit06} find that this morphology is the result of several condensations that are superimposed along the line of sight, whose surfaces are directly illuminated by the star.
The incident radiation field, in units of
the interstellar far-ultraviolet radiation field estimated by Habing\footnote{The Habing  field corresponds to 1.6$\times$10$^{-3}$ erg~s$^{-1}$~cm$^{-2}$ when integrated between 91.2 and 240 nm 
  \citep{hab68}.},  is $G=$2.6$\times$10$^{3}$ \citep{cho88,pil12}. However, higher values also having been reported in the literature \citep{fue99}. Deriving the incident radiation field relies on several assumptions on the geometry and spectral type of the star. For the rest of the paper, we will assume the value of $G=$2.6$\times$10$^{3}$ at the PDR front as in  \citet{cho88} and \citet{pil12}.
  
 NGC\,7023 has been the subject of many studies in the literature, including several spectroscopic studies in the mid- and far-infrared (FIR). The mid- and far-IR spectral regions contain many important cooling lines and hold the fingerprints of the dust emission. \citet{fue00} presented ISO/SWS and
 LWS observations of NGC\,7023 and differentiated between three regions:
 1) the star and the cavity formed by the star, which is filled with a
 low density gas, 2) the edges of the cavity that defines the PDRs, and 3) the molecular gas. Based on their analysis of the
 atomic and molecular lines (\oIa, \oIb, \cII, H$_2$), they find that
 the NW and SW PDRs  have similar excitation conditions, with filaments having a high density of $n\sim$10$^6$ cm$^{-3}$, and a lower density of $n\sim$10$^4$ cm$^{-3}$ in the region between filaments. \citet{wer04} and, later on, \citet{sel07} studied the
variation of the dust features in the mid-IR spectrum of NGC\,7023 using
the Spitzer/IRS and MIPS instruments. They observed variations in the NW and SW
PDRs of the intensity,
widths, and central wavelength of spectral features arising from polycyclic aromatic hydrocarbons (PAHs), as a
function of the distance to the ionising source (HD\,200775).    \citet{wit06} found a good correlation between the optical 
emission from the dust, as traced by the R band (658 nm), with that of ro-vibrational H$_2$. Using Spitzer data, \citet{fle10} studied the H$_2$ and PAHs emission in the PDRs of three reflection nebulae, including NGC\,7023. They showed evidence for PAH de-hydrogenation, which they concluded is suggestive of H$_2$ formation on PAHs. 

Recently, NGC\,7023 has been the target of several programs using
the Herschel Space Observatory \citep{pil10}, vastly improving the spatial resolution of previous far-infrared studies. 
\citet{abe10} combined Spitzer and 
Herschel maps to study the spatial
variations of the dust properties in the east PDR. Using a radiative transfer code they were able to
reproduce the different spatial variations of PAHs, very small grains (VSGs), and large dust grain emission. \citet{job10}
used the high spectral resolution provided by the HIFI instrument to study
the \cII~line at several positions in the nebula, including a central position in the NW PDR.
They found that the emission of \cII~line and PAHs arises from the transition region between 
the atomic and molecular gas. \citet{oka13} studied the photoelectric efficiency from PAHs in a sample of six PDRs, including the East and NW PDRs of NGC7023. Using Spitzer and Herschel/PACS observation, they evaluate the photo-electric effect from PAHs and the cooling \oIc, and \cII~lines in three regions (cavity, interface, and molecular region) and find that regions with high fraction of ionised PAHs have a lower heating efficiency.

In this paper we present  observations from the Photoconductor Array Camera and Spectrometer  \citep[PACS,][]{pog10}  of the \oIa,
\oIb, and \cII~lines in the NW PDR of NGC\,7023. These observations allow us to resolve in detail the global shape of the PDR, although substructures such as the filaments observed at higher spatial resolution \citep[e.g.,][]{lem96, fue96}, are unresolved. This study complements a recent SPIRE spectroscopic study of the molecular gas and dust emission by our group
\citep{koh14}. 
This paper is organised
as follows: the observations and data reduction are
described in Sect.~2. In Sect.~3 the spatial morphology of the lines, their comparison to the emission of CO and H$_2$, and the correlation between the different line ratios as a
 function of the distance to the ionising source,  are  discussed.  In Sect. 4 the cooling budget, and the morphology of the main cooling lines are analysed.  In Sect.~5 the results of the modelling are described. Finally, the conclusions are summarised in the final section.

\section{Observations and Data Reduction}

\begin{figure}[!t]
  \begin{center}
   \includegraphics[width=8.5cm]{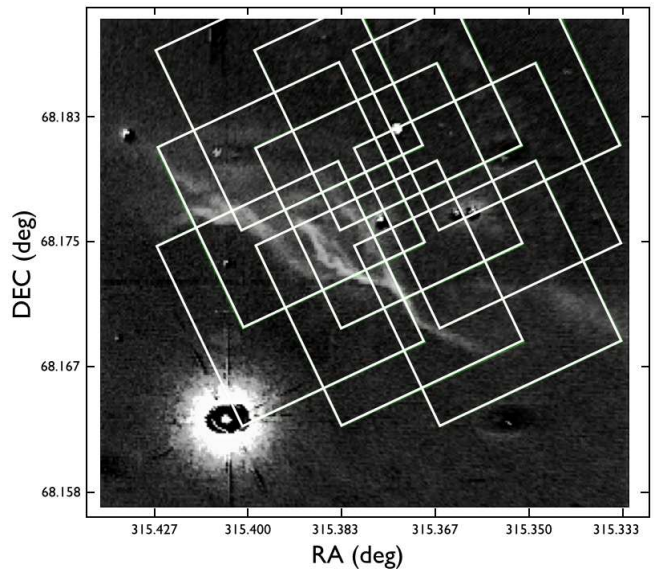}
 \end{center}
  \caption{Overlay of the PACS field of view on a ground based H$_2$ 1-0 S(1) map at 2.12 $\mu$m \citep{lem96}. A 3$\times$3
   raster
    map was performed (see Sect.~2). The map covers an
     area of approximately 110$\arcsec$x110$\arcsec$ and is centred in the
    NW PDR. The emission from the ionising stars (HD\,200775), on the bottom left of the figure,  is saturated. RA and Dec coordinates are in degrees (J2000)}
\end{figure}

The observations were carried out during the science demonstration phase
(SDP) of Herschel on 7 January 2010, and are part of the {\em "Evolution of the
  Interstellar Medium"} guaranteed time key project  \citep[observation
ID=1342191152,][]{abe10}. The
observations were taken using the PACS instrument in the - now decommissioned - wavelength
switching mode. Four fine-structure lines were observed:
\ion{[C}{II]} at 158\,$\mu$m, \ion{[O}{I]} at 63 and 145\,$\mu$m, and
\ion{[N}{II]} at 122\,$\mu$m. The observations were centred on the NW
PDR, with RA and DEC coordinates of 315.375\degr and 60.178\degr respectively
(J2000).

The observing strategy and reduction methods follow that of our
earlier paper on the Orion Bar \citep{ber12}. We thus
summarise the main steps here, and refer to the above paper for
details.

To trace the NW PDR a 3x3 raster map was performed.  The configuration at the time of the
observation is shown in Fig.~1, where the raster map at the epoch of
observation is overlaid on top of an H$_2$ map of the region made with the Canada-France-Hawaii Telescope (CFHT). An
additional map centred in the East PDR (not shown in the figure) will be presented in a future
paper. A minimum exposure configuration of one cycle and repetition
per line was performed. Raster and point line steps of 23.5$\arcsec$
were chosen and result in Nyquist sampling for the lines in the
red channel (\cII , \oIb , \nII). The purpose of the wavelength switching
mode is to cancel out the background by determining a differential
line profile. This means that observations at an off-position are not needed to remove the background in such mode. The present observations of NGC7023 were deemed to have calibration value by the Herschel Science Center, and it was decided to obtain an additional set of observations at the centre position so that the PDR would be observed with a higher S/N. This was achieved by including an off-observation in the astronomical observation request (AOR), to be performed after three raster positions and with zero offset relative to the centre of the map. This resulted in the centre of the map being observed a total of 3 times, with the redundancy resulting  in a higher S/N for that central position, which is centred in the PDR.

The data were processed using version 6.0.3  of the reduction and
analysis package HIPE \citep{ott10}.  Each raster position, or footprint, consists of 5$\times$5 spectral pixels (called spaxels). HIPE produces  a data cube for each footprint, which contains a spectrum for each spaxel. Starting from  level 1 the cubes were
further processed, using proprietary tools, to correct for minor
drifting effects and flux misalignments between scans. 
At this point the cubes are exported into the 
software PACSman \citep{leb12} to measure the line fluxes by fitting
a Gaussian, and create the final map. The uncertainties in the line
fluxes  are small and on average amount to less
than 5\% for the \ion{[C}{II]} and \ion{[O}{I]} lines.
The relative flux accuracy between  spaxels
is 10\%\footnote{From the PACS spectroscopy performance and
  calibration manual. This can be found at
  http://herschel.esac.esa.int/twiki/bin/view/Public/PacsCalibrationWeb}, and for the remaining of the paper we have adopted this
   value or the uncertainty from the fit given by PACSman, whichever is higher.
Note that we do not detect the \nII~line at any position
in the map. 
The final integrated intensity maps of  the \oIa, \oIb, and \cII~lines are shown in Fig.~2. 

\begin{figure*}[!t]
  \begin{center}
 \includegraphics[width=10.cm,angle=90]{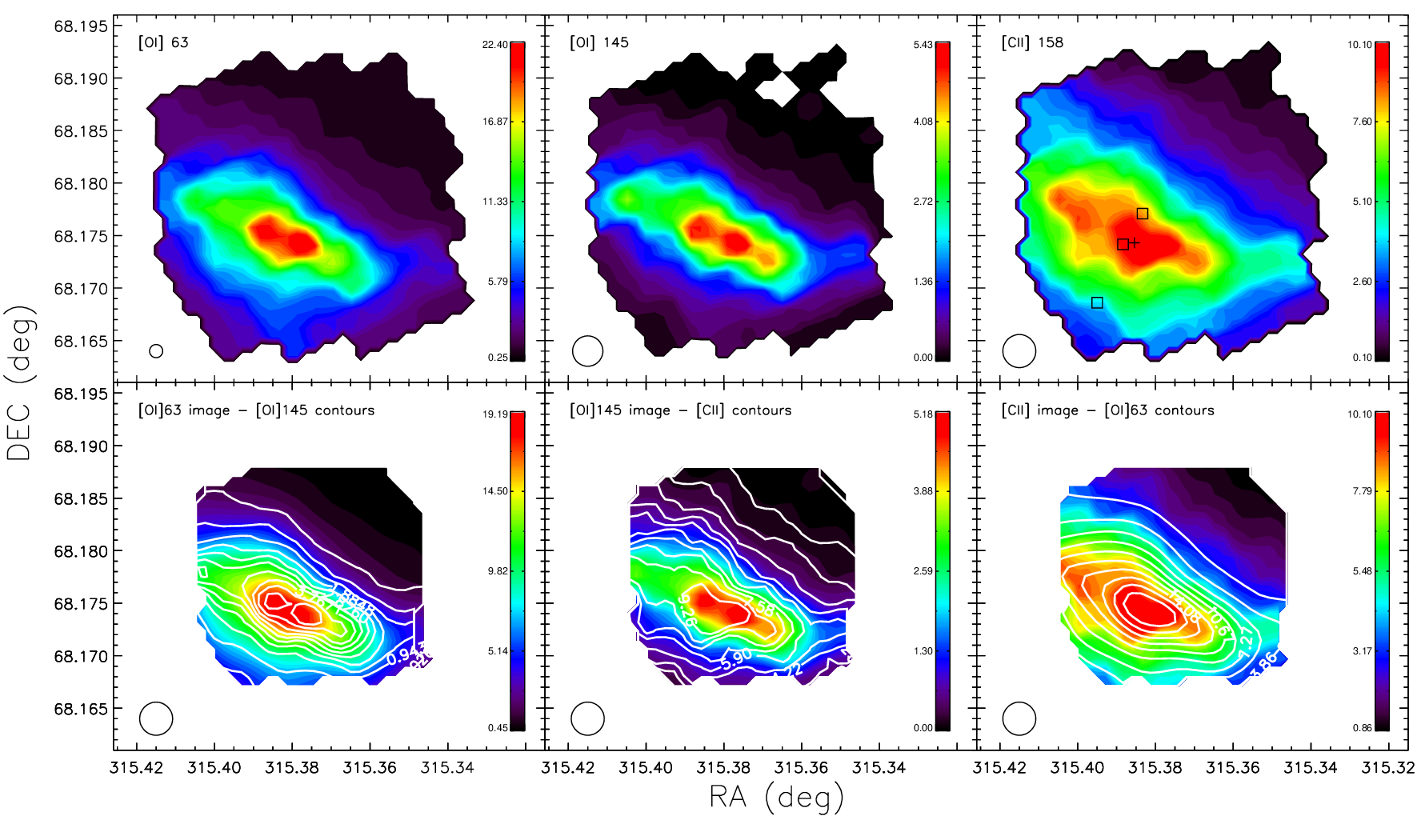}
  \end{center}
  \caption{The upper three figures show the observed  images for the \oIa, \oIb, and \cII~lines. The
   lower panels show combination of lines with contour maps over-plotted in white. The bottom images have been convolved to the 158\,$\mu$m
   (largest) beam, with the beam size illustrated in
   the bottom left of each panel. The squares and plus symbols in the upper right panel indicate the positions used to compare the \cII~ flux to previous measurements in the literature (see Sect. 2). All maps are in flux units of
    10$^{-7}$ W~m$^{-2}$~sr$^{-1}$. The eight highest contours are at 100, 91, 82, 74, 65, 56, 47, and 38\% of the peak emission.}
\end{figure*}

We have compared
our measurements of the \ion{[C}{II]} line with the fluxes measured by
\citet{job10} using the HIFI instrument, and the ISO/LWS measurements by \citet{fue00}. all compared to the largest beam. The position used for comparison (RA$=$315.385$\degr$, DEC$=$68.1743$\degr$) is located in the clump of enhanced emission and is marked at the top right panel of Figure 2 with a plus symbol. All measurements agree within $\sim$10 \%.  \citet{oka13} observed the same lines in the NW PDR. Their observations consists of a single pointing, while our observations perform a detail mapping of the region to show the morphology of these lines, providing a complete coverage of the region and allowing for better calibration. We have compared our fluxes to that of \citet{oka13} in the cavity, interface, and molecular region. These three positions are marked with squares in the top right panel of Figure 2.  All values agree within 29\%, where the \cII~and \oIb~fluxes in the interface, and the \oIa~flux in the molecular region agree within 5\%. The uncertainty in the \citet{oka13} paper is of $\sim$30\%, while in our work is typically $<$10\%, and thus both values are consistent within the uncertainties.

In addition to the atomic lines presented in this paper, we compare  in Section 4 their emission to other relevant cooling species of CO, H$_2$O, CH$^+$, and C$^0$. In particular, we use the CO 4$<$J$_u$$<$13 transitions in the SPIRE range, and the rotational lines of H$_2$ 0-0 S(1), S(2), and S(3) at 17.0, 12.3, and 9.7 $\mu$m from the Spitzer data. We also use the 360 and 609\,$\mu$m lines of C$^0$ (SPIRE). Since we do not detect any CH$^+$ and H$_2$O lines in the SPIRE range, we derive an upper-limit to the transitions of these species that were detected in the Orion Bar \citep{ber12}. These observations were taken from  \cite{pil12} for H$_2$ lines, and \citet{koh14} for the SPIRE lines.

\section{Spatial Distribution}

\subsection{Cooling Lines}

The three top panels in Fig.~2 show the observed distributions of the
\oIc, and \cII~lines. The exciting source, the binary system HD\,200775 (RA$=$315.404$\degr$, DEC$=$68.1633$\degr$), illuminates the
PDR from the lower left side of the figure (just outside the maps, see
also Fig.~3).  For each panel the beams are shown in the lower left
corner and represent sizes of 4.5\arcsec, 8.8\arcsec, and 11\arcsec~from left to right respectively. In this scale, and 
adopting a distance
of 430~pc to NGC\,7023, 10$\arcsec$ corresponds to a physical scale of
0.02~pc. 

The PDR is detected in the
three lines. Their emission arises from the surface of the clouds, where the gas is warm \citep[$\sim$700K-70K,][]{fue99,fle10,koh14}. There is also evidence for condensations of enhanced emission within the PDR, where both \oIc~line maps can resolve two structures (red clumps in the figure). This is the first time  in NGC\,7023 that such structures have been revealed from the emission of  the \cII~and \oIc~lines  within the PDR. These structures or knots are directly illuminated by the star, and coincide with the presence of highly-excited CO lines and dust emission (see next section). These clumps are better delineated in the \oIb~map (Figure 2), which reveals two distinct clumps, and a possibly third fainter clump. The size of the main structures is around 8$\arcsec$ (0.015~pc at the distance to NGC\,7023). While it is possible that these clumps may not be yet be resolved in our observations, we find that these clumps  cover 9\% of the PDR surface area, where in this case we consider the PDR area to be any \oIb~flux greater than 0.9$\times$10$^{-7}$ W~m$^{-2}$~sr$^{-1}$ (red, green, and blue colours in the upper middle-panel of Fig. 2).

The PDR is well delineated in the \ion{[O}{I]}145\,$\mu$m line (Fig.~2). Further
 away from HD\,200775,  the
emission of \oIb~drops to zero. The spatial distribution of the \oIa~line is similar, although the emission is
more extended than that of the \oIb~line. This is expected as the
\oIa~line becomes optically thick at lower column densities than the
\oIb~line, 
and once saturated, its emission will be broader \citep{cau99}. The
\cII~emission is even more extended, since it is easily excited. This was also seen in a similar study by our group
on the Orion Bar \citep{ber12}.  The figure also shows that the morphology of the \cII~line is complex and
structured.
Using Herschel/HIFI high-resolution spectra of the \cII~line, \citet{job10} and Bern\'e et al. (2014, in prep.) detect different 
components with a large distribution of velocities, indicating that besides the main
filaments, other regions are contributing to the \ion{[C}{II]}
emission.
 
In the lower three panels of Fig.~2 we plot different combinations
of lines and their contours convolved to the largest beam size (11\arcsec). It can be seen that both of the \ion{[O}{I]} lines correlate
spatially very well with each other (lower left panel). The middle
bottom panel shows that the \ion{[C}{II]} emission also delineates the
PDR but with a more complex structure than the \oIb~line. The peak emission
is slightly displaced and broader in the \ion{[C}{II]} line than the \oIb. This is better
seen in the last panel where the concentric contours of the
\oIa~line contrasts with the more irregular emission of
\ion{[C}{II]}, including a peak of emission with a  {\em heart-}shape structure.

\begin{figure*}[!ht]
  \begin{center}
  \includegraphics[width=20cm]{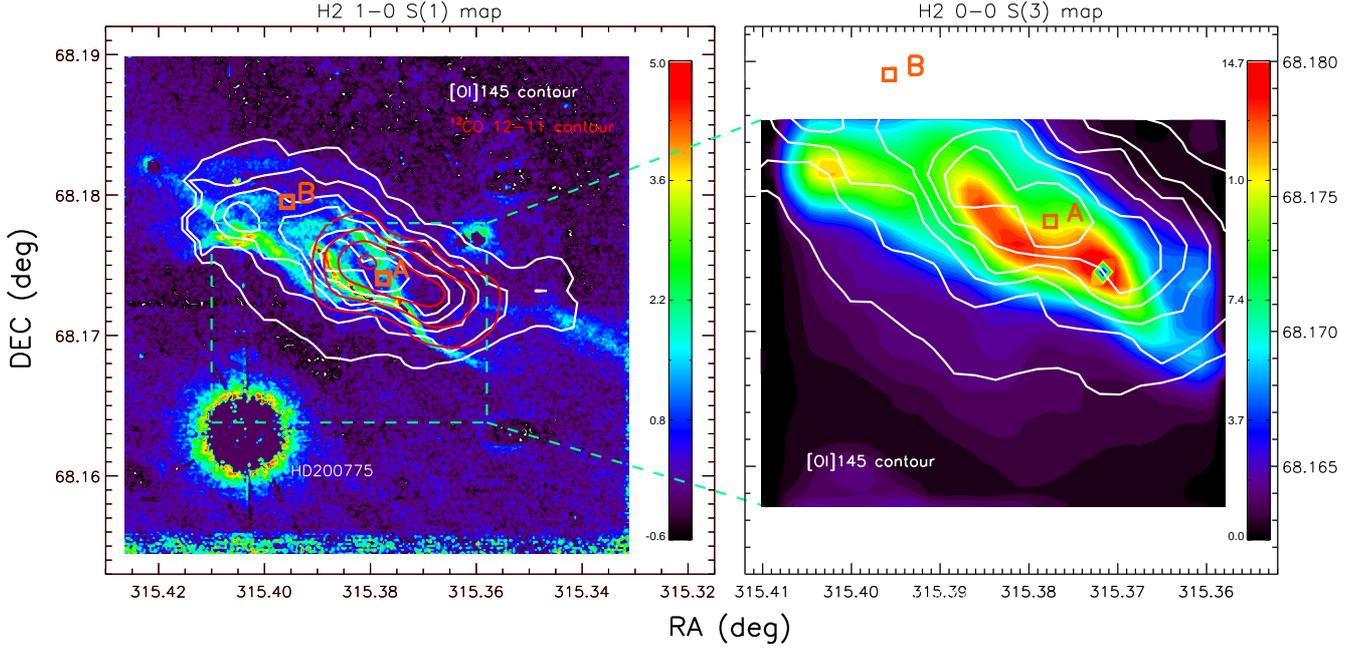}
  \end{center}
  \caption{{\em Left panel:} Spatial correlation of the H$_2$ 1-0 S(1) emission map at 2.12 $\mu$m \citep{lem96}  compared to
    the \ion{[O}{I]} (145$\mu$m) and $^{12}$CO J$=$12-11 \citep{koh14} contour maps in white and red
    respectively. To better illustrate the filaments these images have not been convolved to the
    largest beam (in this case, CO). The two orange symbols (squares) mark the position where the high excitation CO lines are 
    detected (A), and a nearby position with similar conditions but showing no highly excited CO emission (B). These positions are discussed
     in Sect.~3.2. {\em Right panel:}  H$_2$ 0-0 S(3)  map at 9.7$\mu$m convolved to the \ion{[O}{I]} 145$\mu$m beam with its contours in white. In both panels the colour bar intensities are in units of 10$^{-7}$ W~m$^{-2}$~sr$^{-1}$, where the four highest contours have values of  5.4, 4.7, 3.9, and 3.1 for \oIb, and of 0.37, 0.3, 0.2, and 0.1 for the CO.}
\end{figure*}

Finally, as mentioned in the previous section, no \ion{[N}{ii]} line at 122\,$\mu$m was detected in the PACS map. 
The 205\,$\mu$m line was also not detected in the SPIRE maps \citep{koh14}. 
Using HIFI observations that include the ionising stars, the Herschel Warm and Dense ISM program \citep[WADI,][]{oss11} detects 
ionised emission only from a shell close to the stars (private communication), an area just outside our observed map. This could indicate that the  gas in the cavity near 
the PDR is mostly neutral and not ionised, and explains why we do not detect the 122\,$\mu$m~line since our map does not
 include this region close to the ionising stars. Alternatively, if the density is too low in the cavity \citep{fue00} its emission will be too 
 weak to detect in our observations. We have derived an upper-limit to the flux for this line of 1.1$\times$10$^{-7}$ W~m$^{-2}$~sr$^{-1}$.

\subsection{Comparison with CO,  H$_2$, and PAH emission}

   Fig.~3 (left) displays a high-resolution CFHT image of the H$_2$ 1-0 S(1) molecular line at 2.12\,$\mu$m \citep{lem96}, which reveals in detail the
 filamentary structure of the region. Over-plotted in white and red respectively are contours of the \ion{[O}{I]} 145$\mu$m line and one of the excited
 rotational $^{12}$CO line (J$=$12-11) measured in the SPIRE data by \citet{koh14}. We note that the images are not convolved as the
purpose of this panel is to place the emission of \ion{[O}{I]} 145\,$\mu$m line and warm CO relative to the filaments. The right panel of the figure  shows the Spitzer H$_2$ 0-0 S(3) 9.7\,$\mu$m map convolved to the 145\,$\mu$m beam, with white \oIb~contours. The H$_2$ 0-0 S(3) transition follows the same distribution as the 1-0 S(1), but being pure rotational line it peaks slightly behind ($\sim$1.9\arcsec) the 1-0 S(1) transition. These observations show for the first time that the \oIb~emission in NCG\,7023 is displaced with respect to the bright filaments that are seen in H$_2$. The reason for this is that since the H$_2$ emission closely traces  the UV field, its emission peaks at the edge of the PDR, while the \ion{[O}{I]} 145\,$\mu$m is more sensitive to both the gas density and temperature.   \citet{koh14} have recently mapped the emission of high excitation CO lines in the region.  It can be seen that this highly excited CO
emission (left panel in Fig. 3) is located  at the same position as the \ion{[O}{I]} peak. We note that the $^{12}$CO J$=$12-11 line peaks at the 
 same position as the also observed $^{13}$CO J=7-6 line. Similarly, although not shown, these condensations are also detected in {\em Herschel}/SPIRE and IRAM submillimitre maps that trace dust emission \citep{koh14}.
 
It is interesting to explore the origin of this highly excited CO in relation to the atomic cooling line emission. For this purpose, and to guide the discussion in the next section,
we have selected two regions: one coinciding with the CO J$=$12-11
peak (position {\em A}), and a nearby region with no high-J CO emission
({\em B}). We note that low-J emission is detected in positions {\em A} and {\em B} \citep{koh14}. Both positions are labelled in Fig.~3. The \oIb~flux and the  \oIb/\cII~ratio in position A is respectively 2.2 and 1.7 higher than in position {\em B}. Although small, these differences are real (larger than the associated uncertainties in the \oIb~flux and ratio), and  indicate variations in the physical conditions between the 
 two regions considered. Such differences can be achieved 
by changes in the radiation field and/or density of a factor $\sim$3 in standard PDR grid models 
\citep{kau06}.  How can such small differences explain the presence of high-excited CO in position {\em A}?
The emission of highly excited CO can be 
 explained by changes in the gas density if a certain threshold in the conditions is reached. \citet{koh14} derived a 
 density of 5$\times$10$^{4}$--10$^{6}$ cm$^{-3}$ and a temperature ranging between 65--130~K at position {\em A}, while at position {\em B}
  they found a lower  density (10$^{4}$--10$^{5}$ cm$^{-3}$) and {\em T}$=$80--120~K. 
The critical density of the CO lines 
  (10$^5$--10$^{7}$ cm$^{-3}$ for J=4-3 to 13-12 transitions) is higher than the \ion{[O}{I]} lines  ($\sim$10$^5$ cm$^{-3}$). This means that, in the density regime between 10$^{4-6}$ cm$^{-3}$, the CO lines are more sensitive 
  than the oxygen lines to an increase in density which could boost the CO emission while having little effect on the  atomic line emission.   It is thus possible that differences in densities in the two positions could thus explain the presence of highly excited CO in the clumps. 

\begin{figure}[ht]
  \begin{center}
  \includegraphics[width=9cm]{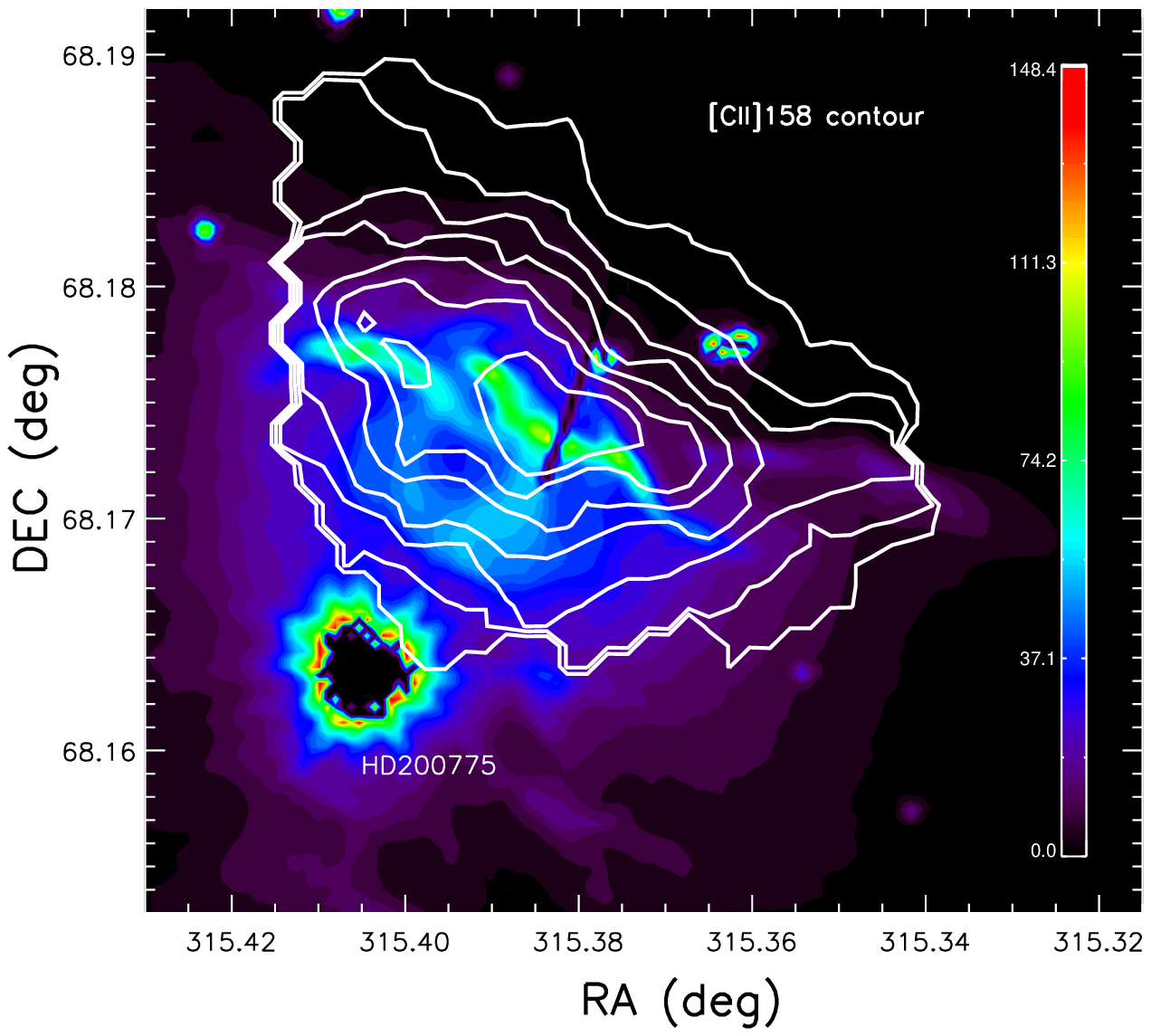}
  \end{center}
  \caption{Comparison of the PAH morphology as traced by the
    IRAC 3.6$\mu$m band intensity in MJy/sr (taken from the Spitzer archive), with  \ion{[C}{II]} 158\,$\mu$m in white contours. In decreasing order the contour levels are at 10.1, 9.0, 7.8, 6.7, 5.6, 4.5, 3.4, and 2.2 $\times$10$^{-7}$ W m$^{-2}$ sr$^{-1}$.}
\end{figure}

Because  of the low ionisation potential required to ionised carbon (11.3 eV), the [CII] 158\,$\mu$m line can arise from both the neutral 
(i.e. PDR) and the ionised medium. To validate its use as star formation indicator, it is interesting to compare the [CII] emission with other known tracers of the
star formation process, like PAHs \citep{pee04}.  In Fig.~4 we make such comparison where we plot the PAH emission traced by the
Spitzer/IRAC 3.6\,$\mu$m band with the \cII~contour map.  Despite the the fact that the \cII~line includes emission from different velocity components,
 it can be seen that the \ion{[C}{II]} line
follows reasonably well the PAH emission, with the peak of \cII~emission
(heart-shaped) coinciding with the PAH filament. Whichever regions contribute to the \cII~emission, these must be (mostly) PDR 
dominated since it traces the PAH emission.

\subsection{Line Correlations and Ratios}

\begin{figure*}
  \begin{center}
   \includegraphics[width=10.cm,angle=90]{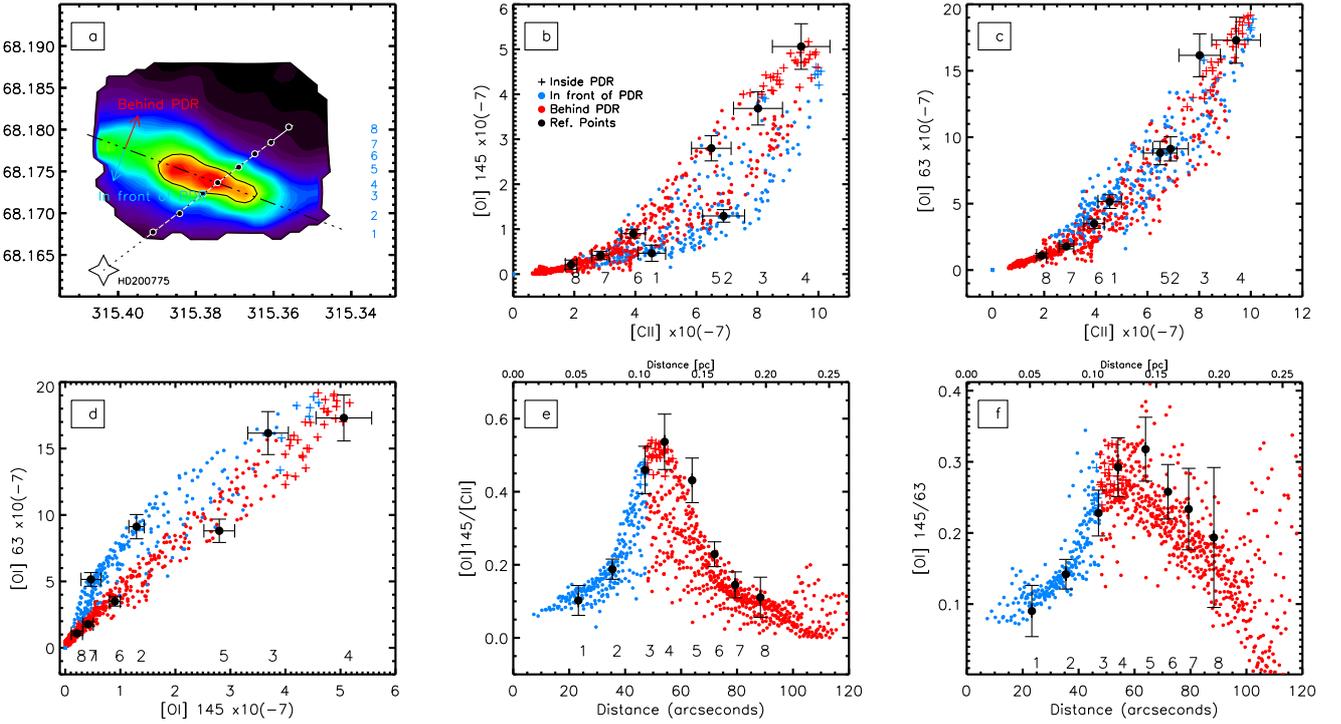}
 \end{center}
  \caption{Intensity plots for different combinations of line fluxes
    (in 10$^{-7}$ W~m$^{-2}$~sr$^{-1}$) and line ratios relative to other lines or distance to HD\,200775.  All fluxes have been 
    convolved to the largest beam (11$\arcsec$). The different regions are indicated in the convolved \oIb~map in the first panel and
    are labelled in the second, where blue represents points in front
    of the PDR,  and red those behind it. The peak emission in the PDR is delineated by the black curve in the first panel (those with \oIb~fluxes
     higher than 75\% of the peak intensity of the line), and are plotted as plus symbols in the subsequent
      panels. The reference points for the adopted cut are numbered and shown in the first panel 
    (Sect. 5), and have been plotted as black dots with error bars in subsequent panels to give an
    indication of the uncertainties.}
\end{figure*}

Different lines and line ratios have been plotted against each other as a function of the distance to the ionising stars HD\,20077 in Fig.~5. 
Panel {\em a} in this figure highlights the different regions we consider. These include
regions in front of and behind the PDR, which in subsequent panels are shown as blue and red dots respectively. 
In the figure the plus symbols are PDR points with
  a flux higher than 75\% of the \oIb~peak emission and are delineated by the black contour. Also shown are selected points
from an adopted cut across the PDR starting from HD\,200775
(labeled numbers). The latter are shown with their error bars to
illustrate the flux uncertainty in different regions of the map. The
variation in the flux intensities across the PDR of the \cII, and \oIc~lines is larger
than a factor 10, which is higher than those observed in the Orion Bar
(factor $<$7) for the same lines \citep{ber12}.

Panels {\em b} and {\em c} show respectively the relation between \ion{[C}{ii]} and
\oIc~lines.  There is a broad correlation in front of and after
the PDR in both plots. The likely reason for the
broad relation is the mixture in the line of sight of the different components 
that give rise to the \cII~line (see Sect. 3). The relation in panel {\em c} is somewhat narrower than
in panel {\em b}. This could be attributed to the fact that the emission of the \oIa~line, being optically thicker than the \oIb, 
is also extended and follows the \cII~emission more closely. 

Panel {\em d} shows a linear correlation between the
\oIc~lines behind the PDR, and a more parabolic relationship in front of it.
In front of the PDR the gas is warmer and optical depth effects become more important. Having an Einstein coefficient five times higher, the  \oIa~line will be more self-absorbed than the \oIb~line, and its overall emission will be more extended. Therefore, in
 that region  \oIb~increases more than the \oIa~line does. 

In the last two panels in the figure ({\em e} and {\em f}), the \oIb/\cII~and \oIb/\oIa~ratios are plotted respectively against the distance to HD\,200775.  As we approach the PDR the density increases
and both line ratios increases rapidly, peaking at around 45$\arcsec$. After
the PDR, the temperature drops and both line ratios decrease. 
However, and because of the
multi-component \cII~emission, there is a broader relation in the
ratio.

\section{Cooling budget}

The mid- and far-IR regions are home to the most important cooling lines in the PDR. Our Herschel observations, together with Spitzer data, have enabled us to estimate the importance of these species to the cooling budget. We have thus derived the contribution of the
[\ion{C}{II}] and [\ion{O}{I}] cooling lines in the PDR, as well as other relevant atomic and molecular
  lines in the mid- and far-IR including; CO, H$_2$, and C$^0$.  H$_2$ and CO are the most abundance molecular species in the ISM and contribute also to the cooling in PDRs \citep{hol99}.  To calculate the budget of these species we consider the 0-0 S(1), S(2), and S(3) rotational lines of H$_2$ at 17.0, 12.3, and 9.7 $\mu$m (Spitzer/IRS), and CO 4$\leq$ J$_u$ $\leq$13 transitions (SPIRE).  In the Orion Bar, C$^0$, CH$^+$ and H$_2$O lines are detected in the PACS and SPIRE spectra and contribute $\sim$1\% of the total cooling budget \citep{hab10, ber12}. This is a small factor to the total cooling budget in the Orion Bar, but we have explored their contribution in NGC\,7023 as well.  We have detected the 370 and 609 \,$\mu$m lines of C$^0$ in the  SPIRE spectrum \citep{koh14}.  We do not detect  CH$^+$ and H$_2$O transition in the SPIRE spectrum of NGC7023 so have derived upper-limits for their far-IR transitions to estimate their contribution.  For this comparison we choose the 
  position of the \ion{[O}{I]} emission peak in the PDR (red knots in Fig. 2), and we convolve all the line emission to the largest beam, in this case the SPIRE CO 
  emission (25\arcsec). 
In Table 1 we quote these values together with those of the Orion Bar \citep{ber12}, which is usually adopted as the prototypical PDR. 

\begin{table}[h]
  \caption[]{Cooling line emission as a percentage at the PDR position and convolving the emission of all species to a 25Ó beam.}
  \begin{center}
  \begin{tabular}{l c c}

  \hline
  \hline
  & NGC\,7023 & Orion Bar$^{\dagger}$  \\ 
  \hline

  \ion{[C}{II]} 158\,$\mu$m    & 19 & 8  \\
  \ion{[O}{I]} 63\,$\mu$m     & 33 & 72  \\ 
  \ion{[O}{I]} 145\,$\mu$m     & 9 & 10  \\ 
  \ion{[C}{II]} + \ion{[O}{I]}   & 61 & 90  \\
   H$_2$                            & 35 & 5    \\
   CO                                 & 4 & 5   \\
  C$^0$, H$_2$O, CH$^+$  & $<$1 & $<$1 \\

  \hline
  \end{tabular}
  \end{center}
  $^{\dagger}$ From \citet{ber12}.\\ 

\end{table}

\begin{figure*}[!t]
  \begin{center}
 \includegraphics[width=12cm,angle=90]{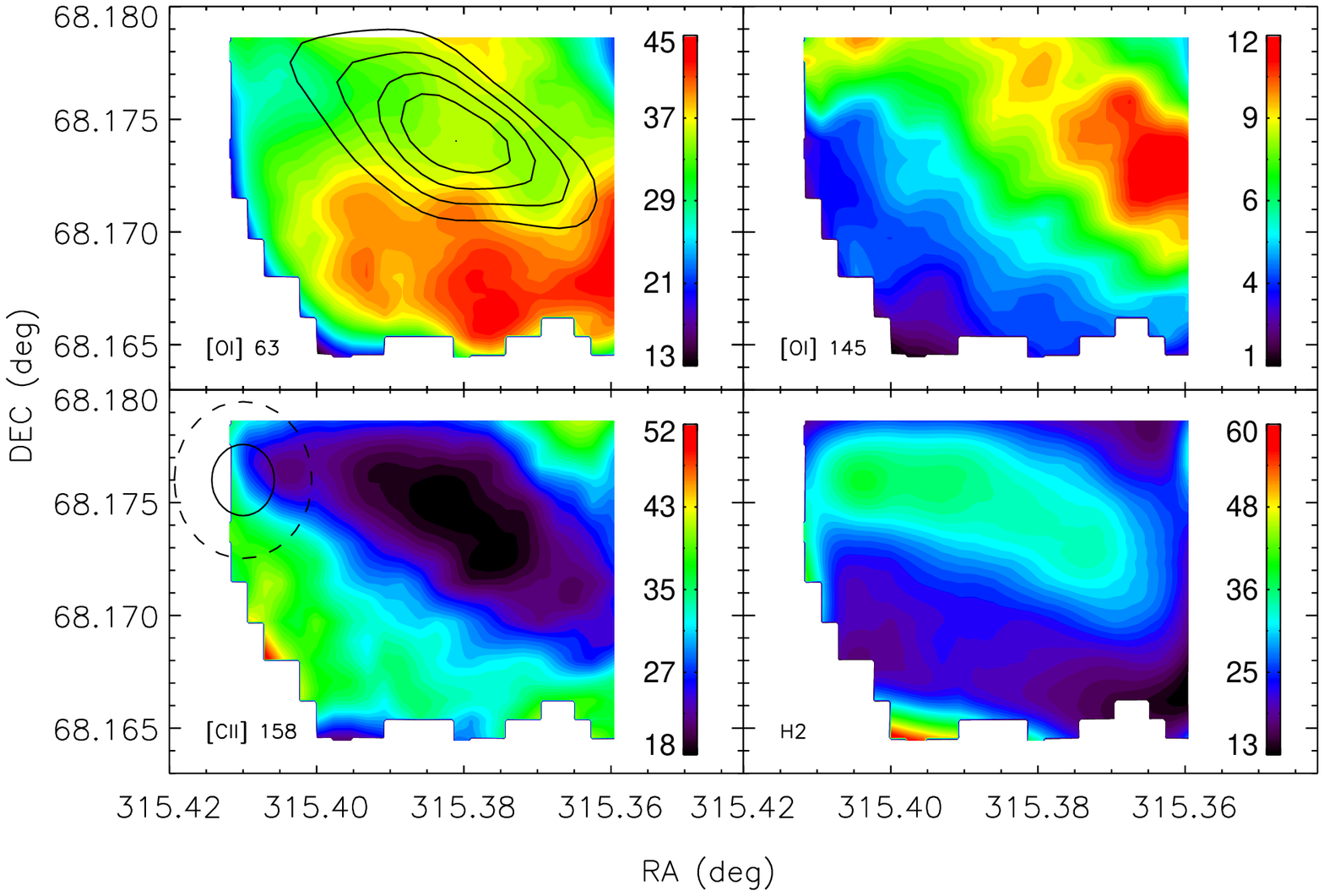}
  \end{center}
  \caption{Relative  cooling budget of the \oIc, \cII~lines, and the 0-0 S(1), S(2), S(3) H$_2$ rotational transitions and v$=$1-0 S(1) in percentage (\%). Adding the  contribution of all four species for a given point in the map amounts to 100\,\%. Other species that can contribute to the cooling are ignored in this figure (see Sect. 4).  In the first panel, and to guide the eye, a contour of the PDR as traced by the \oIa~emission is shown. The images have been convolved to the largest of the beam sizes of these species, in this case the \cII~line, shown as a solid circle in the [CII] panel. In addition, and for comparison, we also show the largest CO beam size (25$\arcsec$) used to calculate the cooling budget in Table 1 as a dashed circle .}
\end{figure*}

Table 1 shows that in the NW-PDR of NGC\,7023 the cooling budget is dominated by the \ion{[C}{II]}+\ion{[O}{I]} (61\%), followed by 
H$_2$ with 35\%, CO (4\%), and C$^0$ contributing less than 1\%. The uncertainty in these percentages is around 20\% of each value. The main contributors to the cooling budget are \oIa~and H$_2$. The overall result is similar to that found in the Orion Bar, where the atomic lines dominate the cooling line emission, with \oIa~being the strongest. 
However, the main difference is the reduced role of H$_2$ in the Orion Bar (5\%) when compared to NGC\,7023. 
As part of the same program, we have studied nine other nearby PDRs and find that, like for NGC\,7023, H$_2$ contributes to about 30\% (Bernard-Salas et al. in prep.). 

In Figure 6 we show the relative contribution by percentage of the \oIc, and \cII~lines, along with that of H$_2$. For H$_2$ we consider the 0-0 rotational transitions S(1), S(2), and S(3) and the vibrational line v$=$1-0 S(1). All maps have been convolved to the \cII~beam. Thus, for a given point in the map, the total cooling of these four species amounts to 100\,\%. Therefore this figure shows which species dominates the cooling in different regions. We ignored the contribution of C$^0$, H$_2$O and CH$^+$, but as we can see from Table 1, their contribution is already negligible in the PDR, where we would expect their contribution to be strongest. If we first look at each map individually, we see that \oIa~contributes more in front of the PDR,  and \cII~outside it (in front and behind). \oIb~and H$_2$ contribute more in the PDR, but with \oIb~contribution being higher in the south-east region (lower-right). The \oIb~line is more sensitive to the gas density than the H$_2$ lines, and thus contributes more where the clumps are located. Reading the percentage from the colour bars, we can also compare the relative contribution amongst each other. In general, \oIa~is the dominant  contributor over the entire region ($>$30\%). However, \cII~ is also an important coolant outside of the PDR (contributing above 30\%), as is H$_2$ in the PDR ($>$35\%). On the other hand, the contribution of \oIb~does not exceed 12\% in the PDR and is lower than 5\% in the rest of the region. This figure delineates the importance of \oIa, \cII, and H$_2$ as cooling agents as a function of the region.

\section{Modelling}

\begin{figure*}
  \begin{center}
   \includegraphics[width=7.4cm,angle=90]{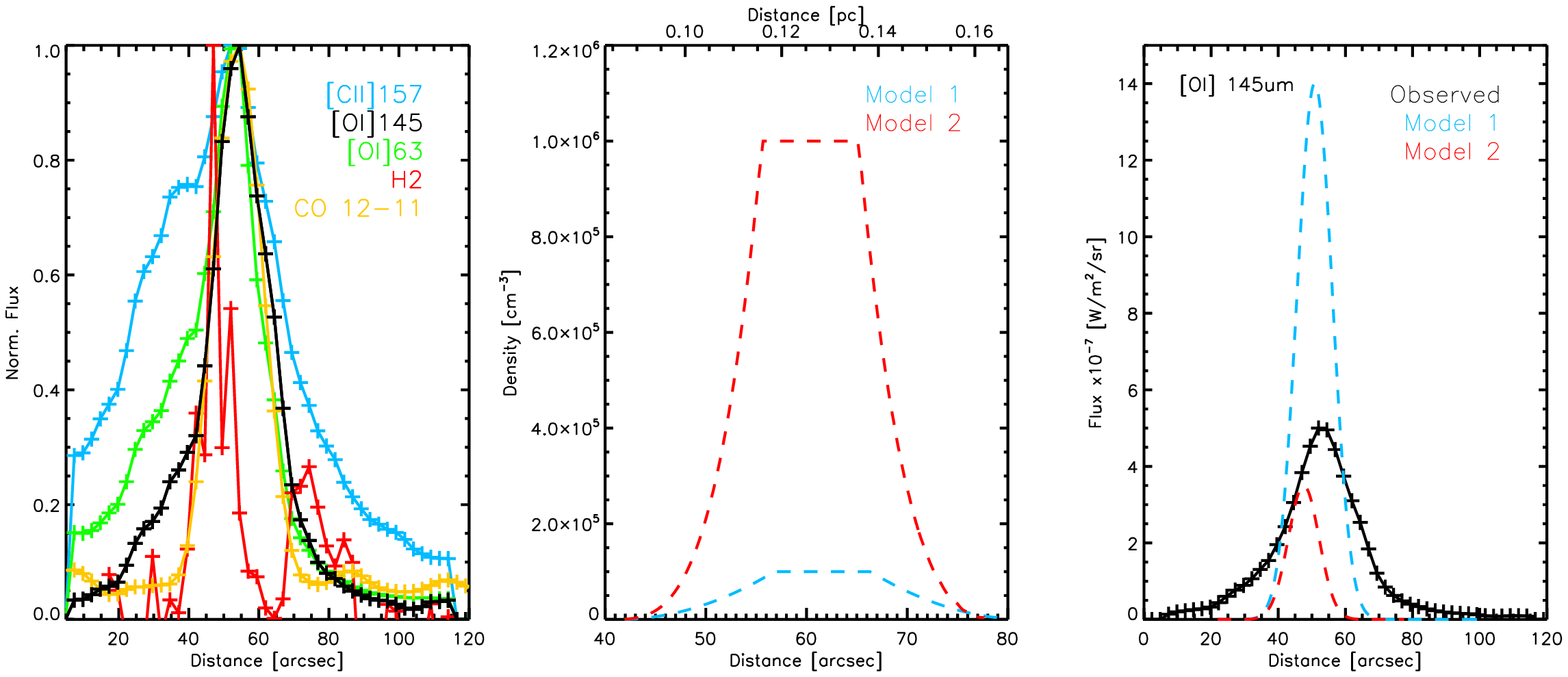}
  \end{center}
  \caption{{\em Left panel:} Observed non-convolved profile of the atomic and molecular lines as a function of the distance to 
  HD\,200775 along the cut shown in Fig.~5 and described in Sect. 4. {\em Middle panel:} Density profiles at two different maximum densities used by \citet{koh14} in their analysis of the molecular emission.  Model 1 has a peak gas density of 10$^5$ and model 2 of 10$^6$ cm$^{-3}$. {\em Right panel:} Observed profile emission of the \oIb~line compared to the
   convolved predicted emission  by Meudon PDR models.}
\end{figure*}

In this section our goal is to explore whether previously reported parameters used to model the dust emission in the PDR (density profile, geometrical parameters), and CO lines (gas physical conditions), can also  reproduce the \oIb~emission. This is important in our quest to find a model that can self-consistently reproduce the dust, molecular, and atomic PDR components. Using Herschel photometry and spectroscopy, the dust emission and molecular content (CO) of the NW PDR has been recently modelled  by \citet{ara12} and  \citet{koh14}.  In our observations the \cII~line is a mixture of many components, the
emission  of the \oIa~line is optically thick, and the \nII~was not 
detected. Thus, attempting to  model the atomic component from our observations will serve to provide basic physical conditions
of the PDR based on just fitting the \oIb~line. This is why we prefer instead to test whether previously and more detailed modelling studies can also explain the \oIb~emission.

In order to model the emission of the line we have selected a cut originating from the ionising stars and
 intersecting the NW PDR at a point of peak dust emission at 250\,$\mu$m. This is the same cut used in  \citep{koh14} for the molecular 
line and dust emission. The observed emission profiles for the three  lines, together with that of the H$_2$ 1-0 S(1)
 and CO J$=$12-11 lines, are shown in  Fig.~7 (left panel). These profiles are shown at their natural 
 resolution to maximise the spatial information we can obtain from them. It can be seen that \oIc, \cII, 
and CO J$=$12-11 peak at the same position. Having the highest spatial resolution, the H$_2$ profile resolves two 
filaments (see Fig. 3), with the densest filament peaking immediately in front of the fine-structure and CO lines relative to the external illumination. The \cII~line has the broadest profile, followed by \oIa.

We have modelled the emission of the \oIb~line with the Meudon PDR radiative transfer code \citep[version 1.4.4, ][]{lep06,lebo12}. This code assumes a plane-parallel slab of gas and dust that is illuminated by an incident radiation field, and resolves iteratively the radiative transfer and thermal and chemical balance to compute the atomic and molecular emission in the cloud.  As in the papers listed above, we have assumed an incident radiation field of {\em G}$\sim$2,600 on the NW PDR at 42'' from the star \citep{cho88,pil12}.  The RADEX analysis of the CO lines \citep{koh14},  and the modelling of the dust emission  are compatible with  two maximum gas densities of 10$^5$ and 10$^6$ cm$^{-3}$. Thus, for our comparison we consider both possibilities.  These studies derive a PDR length of 0.2~pc for a gas density 10$^5$ cm$^{-3}$ and 0.03~pc for a density of 10$^6$ cm$^{-3}$, giving both a total column density of about 10$^{22}$\,cm$^{-2}$, and we adopt these values in our study.  We assume the density profile derived from the dust analysis and adjusted to start at 0.088 pc (42$\arcsec$) from HD 200775. The density profiles for the two gas densities we consider are shown in the middle panel of Fig. 7. We hereafter refer to these models as model 1 (n$=$10$^5$ cm$^{-3}$) and and model 2 (n$=$10$^6$ cm$^{-3}$), respectively.
 
Before we proceed,  it is interesting to investigate whether the depth of the PDR deduced from the previous studies \citep{ara12, koh14} is compatible with the width and length of the PDR we derive from the observations (spatial extent of the \oIb~line).   In the middle-top panel of Fig. 2, the PDR has an approximate width of  20$\arcsec$ and a length of  60$\arcsec$ (as traced by the green colours in the figure), which corresponds to 0.04 and 0.12 pc respectively. From model 1 we find that the depth is about 2 times the observed length, resulting in a flat extended geometry. 
For model 2 the the depth compares very well to the projected width resulting in a cylindric geometry.
 
The predicted \oIb~intensities for models 1 and 2 are shown in the righthand panel of Fig. 7, where the line emissivities have been convolved to the PACS beam at 145\,$\mu$m. For this comparison we consider the conditions of the models in the region where the predicted unconvolved flux is greater than half the maximum flux (48$\arcsec$--52$\arcsec$ from the binary system).  In this region model 1 gives $n=$10$^4$--8$\times$10$^4$ cm$^{-3}$ and $T=$90--530~K. Model 2 gives $n=$2$\times$10$^4$--2$\times$10$^5$ cm$^{-3}$ and $T=$90--660~K. In both models most of the emission comes from a region with a visual extinction $A_V=$ 0.07--1.2. 
With a 10\% uncertainty in the observed line, none of the models reproduces the \oIb~peak intensity. Note that we did not attempt to fit the line, but this additional constraint may serve to discriminate between the two models. Model 1 predicts an intensity nearly three times higher than the observations,  while model 2 underestimate the emission by just 30\%.
The  maximum gas temperature in both models  is higher than the gas temperature derived by \citet{koh14} when modelling the CO J$=$13-12 to J$=$6-5 lines  with RADEX (140~K).
The  higher temperature we find is expected  since models predict that the \oIb~line originates from a slightly warmer layer than that of the intermediate J CO excited lines.

The fact that we find a model that can reproduce the dust emission, as well as roughly matching the spatial profile of the \oIb~line is an encouraging step towards a self consistent model that explains both the gas and dust. However, while model 2 does not reproduce exactly the peak and width of the \ion{[O}{I]} emission, we must bear in mind that we are assuming a simple model with a simple geometry. The H$_2$ observations reveal a more complex structure with several filaments. Also the density profile we use reflects a smooth PDR, while our own observations clearly reveal the presence of dense structure inside of the PDR. Our cut is also arbitrary. These complex structures will affect the \ion{[O}{I]} emission profile (width, peak position). Given these caveats, it is therefore unsurprising that we do not have a perfect fit. In fact, in order to explain the intensity peak of the excited CO emission,  a steeper density gradient at the PDR edge over a small scale is needed,  to have both a large amount of warm and dense gas. The CO lines can be reproduced using a high pressure model ($P=$10$^8$ K~cm$^{-3}$) and relatively small PDR width and depth of 2--3$\arcsec$ (Joblin et al. in prep.). Such a model can also predict the maximum peak intensity of the \oIb~line. A lower pressure is however needed to reproduce the full extent of the PDR that we have presented in Fig.2.

Finally, and since the Meudon code does not takes into account the ionised region, we have modelled the emission of the \ion{[N}{ii]}122 and 205 $\mu$m lines using the radiative transfer code Cloudy \citep{fer98}. We use the same 
 density profile,  peak density, and PDR length as in model\footnote{Using the parameters of model 1 does not change the conclusion.}  2. For the radiation field, we assume a Kurucz spectrum \citep{kur93} with {\em T}$_{\rm eff}=$15,000 K corresponding 
 to the spectral type (B3V-B5V) of the illuminating binary star \citep{ale08} but attenuated in order to have
  an incident radiation field of {\em G}$_0$$\sim$2,600 at the edge of the PDR \citep{pil12}.  While there is a cavity, the ionising sources are young and likely to have some material around the star. This attenuation of the radiation field is likely caused by the envelope of the star and the dilution effect. We also adopt ISM abundances \citep{sav96, mey97, mey98}.  Between the star and the PDR 
  there is a cavity. The density profile is adjusted to start at 0.023~pc (11$\arcsec$) from HD\,200775. 
    The intensity of the nitrogen lines at 122 and 205\,$\mu$m are sensitive to the electron
density in the ionised region, and on the incident radiation field, but we do not detect either line. The upper 
limits we have derived for these lines (1.1 $\times$10$^{-7}$ and 10$^{-9}$ W~m$^{-2}$~sr$^{-1}$ respectively) are compatible with two scenarios for the cavity, one with an ionised 
region and a density {\em n}$<$100 cm$^{-3}$, and another one with no ionised region and a density of 
500 cm$^{-3}$.

\section{Summary and Conclusions}

We have presented {\em Herschel}/PACS spatially resolved observations of the 
\ion{[C}{II]}158\,$\mu$m, \ion{[O}{I]}\,63$\mu$m and 145\,$\mu$m 
lines of the NW PDR in NGC\,7023. This has enabled us to study the
emission of these cooling lines in relation to the morphology of the
region. We summarise here the major findings:

\begin{enumerate}

\item The emission of the atomic cooling lines trace the cloud surface that is directly illuminated by the binary system HD\,200775.  Here the gas is warmer, and these lines are  associated with the filamentary structure revealed in higher-resolution H$_2$ and PAH maps. In the PDR, the peak of cooling line emission corresponds to the presence of condensations that are seen from high level rotational CO lines, and Herschel and submillimetre emission of dust.

\item By comparing the role of different coolants we find that  the \oIc ~and \cII~lines account for 61\% of the cooling in the PDR, with 
\oIa~contributing to 33\% 
of the emission. H$_2$ also contributes significantly with 35\%, CO with 4\%, and other atoms and molecules (C$^0$, H$_2$O, CH$^+$) 
contributing less 
than 1 percent.  Looking at the relative contribution of the main cooling agents over the region, we find that while the \oIa~ dominates the cooling, H$_2$ also contributes significantly in the PDR, and \cII~is also important in front of and behind the PDR.  This highlights the importance of \oIa~and H$_2$ as a coolant in the PDR, and of \cII~in lower excited regions. It is interesting to notice that while in the Orion Bar the atomic lines dominate the cooling, H$_2$ contributes to only 5\%.

\item  The \oIc~maps spatially resolve these condensation into two distinct structures. Furthermore,  {\em Herschel}'s high-angular resolution shows that the \oIb~emission peaks slightly farther away from the ionising stars than the H$_2$ emission which traces the edge of the PDR. This is expected since \oIb~is more sensitive to the gas density. 

\item \oIc ~and \cII~peak at the same position. The \cII~line is more extended than the \oIc~lines because \cII~is easily excited and  presents a more complex structure. This is consistent with HIFI observations of the region by \citep{job10} which reveal different velocity components for the \cII ~emission, indicating that different regions are contributing to the emission of this line. 

\item The emission of highly excited $^{12}$CO J$=$12-11 correlates with the peak of emission of the
  cooling lines. We find that emission from the atomic cooling lines and line ratios at this position differ by a factor of two compared with the conditions of a nearby position with no 
  CO J=12-11 emission. This difference in the cooling line emission indicates a change in the physical conditions (density varying from 10$^4$ to 10$^{5-6}$ cm$^{-3}$). In this density range, the CO lines are more sensitive  than the atomic lines  and could explain the presence of warm CO in this region.

\item Using a density profile derived to reproduce the dust emission and physical parameters from the analysis of CO lines,  we have modelled the emission of the \oIb~line using two different peak gas densities using the Meudon code.  The best model predicts the \oIb~emission to within 30\%, of the intensity at the peak of the emission, occurring at $A_{\rm V}=$0.07-1.2. In this region the conditions are
$n=$1.7$\times$10$^4$--1.8$\times$10$^5$ cm$^{-3}$, $T=$90--660~K. The PDR depth along the line of sight needed to reproduce the absolute \oIb~intensity is 0.03~pc, which is comparable to the observed width of the PDR in our map. 

\item We did not detect the ionised line of \ion{[N}{ii]} at 122\,$\mu$m. Our Cloudy model indicate that the ionised shell may have a  low density, 
in which case,  the \ion{[N}{ii]} 122\,$\mu$m line is too weak to be detected in our observations. Upper-limits to this line and the \ion{[N}{ii]} at 205\,$\mu$m line are consistent with two scenarios for the cavity, one with an ionised 
region and a density {\em n}$<$100 cm$^{-3}$, and one with no ionised region and a density of 
500 cm$^{-3}$.  This is supported by supporting HIFI observations which reveal that the \ion{[N}{ii]} at 205\,$\mu$m  line is only  emitting in a shell close to the star, and indicates  the cavity 
between the star and the PDR is mostly filled with non-ionised gas.

\end{enumerate}

The morphology of the \oIc~and \cII~lines give an unprecedented view into the variation of the physical conditions, energetics, and cooling across the NW PDR.  The clumps of emission we observe provide a means to understand how stellar radiation interacts with the gas, and in the future it will be interesting to model whether such clumps will be photo-evaporated or form stars. Given the complexity of the region, and that we have assumed a simple planar model to reproduce the \oIb~using parameters derived from modelling the dust and CO emission, this study offers an encouraging step forward in producing a self-consistent model that can explain the atomic, molecular, and dust emission. Moreover, spatially resolved studies like this can help us calibrate the use of the  \cII~and \ion{[O}{I]} lines to trace star formation\citep{sta91,mei07}, and understand the origin of the so called [CII] deficit  \citep{luh03} in most luminous galaxies. Because of its brightness, the \cII~line provides an unmatched opportunity for redshift determinations and source diagnostics of far distant 
systems, especially with ALMA.

\begin{acknowledgements}

  We thank an anonymous referee for useful comments and suggestions. JBS wishes to acknowledge the support from a
  Marie Curie Intra-European Fellowship within the 7th European
  Community Framework Program under project number 272820.  MHDvdW is supported by CSA and NSERC. We thank C. Joblin for useful comments and discussion. HCSS,
  HSpot, and HIPE are joint developments by the Herschel Science
  Ground Segment Consortium, consisting of ESA, the NASA Herschel
  Science Center, and the HIFI, PACS and SPIRE consortia. PACS has been developed by a consortium of institutes led by MPE (Germany) and including UVIE (Austria); KU Leuven, CSL, IMEC (Belgium); CEA, LAM (France); MPIA (Germany); INAF-IFSI/OAA/OAP/OAT, LENS, SISSA (Italy); IAC (Spain). This development has been supported by the funding agencies BMVIT (Austria), ESA-PRODEX (Belgium), CEA/CNES (France), DLR (Germany), ASI/INAF (Italy), and CICYT/MCYT (Spain).

\end{acknowledgements}


\end{document}